\documentclass[11pt,a4paper]{article}
\pdfoutput=1
\usepackage{authblk}
\usepackage[english]{babel}  
\usepackage{amssymb,amsmath}    
\usepackage{fancyhdr}       
\usepackage[section]{placeins} 
\usepackage{psfrag} 
\usepackage{verbatim} 
\usepackage{epsfig}
\usepackage{enumerate}   
\usepackage{graphicx} 
\usepackage{caption}
\usepackage[ps2pdf]{hyperref}       
\usepackage{graphics}
\usepackage[sort&compress,numbers]{natbib}
\usepackage{anysize}
\marginsize{2.5cm}{2.5cm}{1cm}{1cm} 

\newcommand{\bmb}{\begin{bmatrix}} 
\newcommand{\bme}{\end{bmatrix}}   
\newcommand{\ppmb}{\begin{pmatrix}} 
\newcommand{\ppme}{\end{pmatrix}}   
\newcommand{\equb}{\begin{equation}} 
\newcommand{\eque}{\end{equation}} 
\newcommand{\equab}{\begin{eqnarray}} 
\newcommand{\equae}{\end{eqnarray}} 

\begin{document}

\title{Neutrino mixing and the Frobenius group $T_{13}$}
\author[1,2,\footnote{Corresponding author\\  Email: christi.hartmann@gmail.com}]{Christine Hartmann}
\author[1]{A. Zee}
\affil[1]{Kavli Institute for Theoretical Physics, University of California, Santa Barbara CA 93106, USA}
\affil[2]{Niels Bohr International Academy and Discovery Center, Niels Bohr Institute, DK-2100 Copenhagen, Denmark}
\date{\today}


\maketitle

$\bold{Abstract}$
\\
\\
We show that the Frobenius group $T_{13} = Z_{13} \rtimes Z_3$ is a suitable family symmetry group, to study neutrino oscillations. Our approach is to catalog all possibilities within an effective field theory approach, assuming only SU(2)xU(1), supplemented by family symmetry. We will use tribimaximal mixing as a guide to place a constraint on the otherwise various possibilities. This leads to an exact fit between the neutrino and charged lepton sector. Such a fit has not been achieved with any other group so far. The results of this paper may then be useful in future studies on the compatibility of this Frobenius group with other models and mechanisms.

\section{Introduction}

According to the Standard Model, neutrinos do not have a mass. However, experiments have shown that neutrinos oscillate, and in order to do so, there must be a difference in their masses, so that at least two of them are massive. Evidence that neutrinos undergo oscillations was first suggested in the 1960's by R. Davies where a lack of observed solar electron neutrinos and atmospheric muon neutrinos compared to the expected amount was measured. Several experiments have since then confirmed this deficit, and in 1998, the existence of neutrino masses was manifested from observed atmospheric neutrino oscillations \cite{Kamiokande}.

The existence of a family symmetry group $F$ (also called horizontal symmetry) was proposed long ago \cite{Wilczek}. Under $F$, the 3 quark and lepton generations transform into each other. Considerable effort has been put into deriving the mixing and mass matrices for neutrinos, using non-abelian finite groups. The tetrahedral group $A_4$ has been studied extensively \cite{Softly,Examples,Models,Babu,Tetrahedral,Altarelli}. So far, this group seems to present the best fit to describe neutrino oscillations. However, other candidates can be found. The Frobenius groups, as sugroups of SU(3), seem to provide interesting family symmetries. We will focus on the Frobenius group $T_{13}=Z_{13} \rtimes Z_3$ in this paper. As usual, the charged lepton fields (left and right handed), the neutrino fields, and various Higgs fields are to be assigned to various representations of $T_{13}$. Interestingly, $T_{13}$ has 4 different 3-dimensional representations, which thus allow for a host of possibilities for model building. It has been shown in \cite{Parattu} that $T_{13}$ is suitable to produce tribimaximal mixing. Instead of making a more-or-less random choice, we catalog and study the various possibilities, using the tribimaximal mixing matrix as a guide. We have subsequently learned that the group $T_{13}$ has also been discussed in \cite{Kajiyama}. This paper makes a particular choice of representations and also proposes a theory of decaying dark matter based on this group. We also note in this context that another Frobenius group, $T_{7}=Z_7 \rtimes Z_3$, has been discussed in \cite{Luhn} and \cite{Cao} in connection with neutrino mixing.

We will start by a brief review on neutrino mixing and then discuss the group theory of the Frobenius group $T_{13}$. After describing how to extract the charged lepton and neutrino mass matrices from the Lagrangian, we will proceed to the model building, where we assign different representations to the respective fields involved, and work out some interesting cases. We show, that this Frobenius group, with tribimaximal mixing as a guideline, has an even better fit between the charged lepton sector and the neutrino sector, than the tetrahedral group.

\section{Neutrino mixing}

Neutrino oscillations require a mixing matrix that transforms the mass eigenstates into the flavor eigenstates as follows:

\equb \left( \begin{array}{ccc}
\nu_e \\ \nu_{\mu} \\
\nu_{\tau} \end{array} \right)= V \left( \begin{array}{ccc}
\nu_1 \\ \nu_2 \\
\nu_{3} \end{array} \right) \label{mix}\eque

This mixing matrix is given by:

\equb \nonumber
V =  \left( \begin{array}{ccc}
e^{i\kappa_1} & 0 & 0 \\
0 & e^{i\kappa_2} & 0 \\
0 & 0 & e^{i\kappa_3}  \end{array} \right) U \left( \begin{array}{ccc}
e^{i\Phi_1} & 0 & 0 \\
0 & e^{i\Phi_2} & 0 \\
0 & 0 & 1  \end{array} \right) \eque
where
\equb U =  \left( \begin{array}{ccc}
c_{12}c_{13} & s_{12}c_{13}  & s_{13}e^{-i\delta} \\
-s_{12}c_{23} - c_{12}s_{23}s_{13}e^{i \delta} & c_{12}c_{23} - s_{12}s_{23}s_{13}e^{i\delta} & s_{23}c_{13}\\
s_{12}s_{23} - c_{12}c_{23}s_{13}e^{i\delta} & -c_{12}s_{23} - s_{12}c_{23}s_{13}e^{i\delta} & c_{23}c_{13} \end{array} \right) \eque
The three $\kappa$ phases can be absorbed into the other phases, by rephasing the neutrino field \cite{Phenomenology}. For Majorana neutrinos, besides the Dirac phase $\delta$, there are two extra Majorana phases, $\Phi_1$ and $\Phi_2$. These are relative phases among the Majorana masses. The Majorana phases are not observable in neutrino oscillations, as is well known.

Experimental measurements are well described by the tribimaximal mixing matrix \cite{Scott}, which we will use as a guide for model building in this paper:

\equb U_{TB} = \left( \begin{array}{ccc}
\frac{2}{\sqrt{6}} & \frac{1}{\sqrt{3}} & 0 \\
-\frac{1}{\sqrt{6}} & \frac{1}{\sqrt{3}} & \frac{1}{\sqrt{2}} \\
-\frac{1}{\sqrt{6}} & \frac{1}{\sqrt{3}} & -\frac{1}{\sqrt{2}} \end{array} \right) \eque

\section{The family group}

The Frobenius group $T_{13}$ is a subgroup of SU(3) generated by 2 elements of SU(3), $a$ and $b$, such that $a^{13} = I$, $b^3 = I$ and $ba = a^3b$. All elements $g$ of $T_{13}$ can be written as $g = a^mb^n$ with $0 \leq m \leq 12$ and $0 \leq n \leq 2$. Thus, the group has $13 \cdot 3 = 39$ elements and 7 conjugacy classes given by:

\equb \nonumber C_1: \{e\} \eque
\equb \nonumber C_{13}^{(1)}: \{b, ba, ba^2, \,\, ...  \,\, , ba^{11}, ba^{12}\} \eque
\equb \nonumber C_{13}^{(2)}: \{b^2, b^2a, b^2a^2, \,\, ... \,\, b^2a^{11}, b^2a^{12}\} \eque
\equb \nonumber C_{3_1}: \{a, a^3, a^9 \} \eque
\equb \nonumber C_{\bar{3}_{1}}: \{a^4, a^{10}, a^{12} \} \eque
 \equb \nonumber C_{3_2}: \{a^2, a^5, a^6 \} \eque
\equb C_{\bar{3}_{2}}: \{a^7, a^{8}, a^{11} \} \label{classes} \eque
From $\Sigma_i d_i^2 = 39$, where $d_i$ are the dimensions of the 7 irreducible representations (IR) and from $\Sigma_i m_i = 7$, where $m_i$ is the number of IR with dimension $d_i$, we see that the group consists of three 1 dimensional IR's $\bold{1}, \bold{1'}$ and $\bold{\bar{1}'}$ and four 3 dimensional IR's, $\bold{3_1}, \bold{\bar{3}_1}, \bold{3_2}$ and $\bold{\bar{3}_2}$. Explicitly, for $\bold{1'}$ and $\bold{\bar{1}'}$, $a$ is represented respectively by $\omega$ and $\omega^2$ and $b$ by 1, where $\omega = e^{\frac{2\pi i}{3}}$ is the cube root of the identity. In the four 3 dimensional IR's, $a$ is represented by
\equb a_1 = \left( \begin{array}{ccc}
\rho & 0 & 0 \\
0 & \rho^{3} & 0 \\
0 & 0 & \rho^{9}  \end{array} \right) \eque
\equb a_2 = \left( \begin{array}{ccc}
\rho^2 & 0 & 0 \\
0 & \rho^{6} & 0 \\
0 & 0 & \rho^{5}  \end{array} \right) \label{a2}\eque
and the corresponding complex conjugates:

\equb a_1^* = \left( \begin{array}{ccc}
\rho^{12} & 0 & 0 \\
0 & \rho^{10} & 0 \\
0 & 0 & \rho^{4}  \end{array} \right) \label{complex} \eque
\equb a_2^* = \left( \begin{array}{ccc}
\rho^{11} & 0 & 0 \\
0 & \rho^{7} & 0 \\
0 & 0 & \rho^{8}  \end{array} \right) \eque
and $b$ by
\equb b = \left(\begin{array}{ccc} 0 & 1 & 0 \\
0& 0 & 1\\
1 & 0 & 0 \end{array} \right) \eque
where $\rho=e^{\frac{2\pi i}{13}}$ is the $13^{th}$ root of the identity. 

Another way of defining the group $T_{13} = Z_{13} \rtimes Z_3$ is to demand that for any element $b$ of $Z_3$ on any element $a$ of $Z_{13}$ \cite{Ramond}, we have:
\equb bab^{-1} = a^r \eque
with $r$ some integer. Letting the element of $Z_3$ act on the element of $Z_{13}$ three times we get:

\equb b^3ab^{-3} = a^{r^3} = a \eque
So we have $r^3 = 1 + 13n$, with $n$ some integer. Clearly, $r = 3$ and $n = 2$ solves this equation, so that:

\equb bab^{-1} = a^3 \eque
($r \neq 1$ is indeed what the symbol $\rtimes$ means). The Frobenius group $T_{13}$ thereby has the presentation:

\equb \langle a, b\mid a^{13} = b^3 = I, bab^{-1} = a^3 \rangle \eque
\\

To find the Kronecker products  and the Clebsch-Gordan decompositions of the IR's, it is convenient to use the fact, that the Frobenius group is a subgroup of SU(3), and exploit the well-known properties of SU(3). Denote the 3-component vector of the defining representation $\bold{3}$ by $\psi^{\mu}$, with $\mu = 1,2,3$, and the vector of the conjugate representation $\bold{\bar{3}}$ by $\psi_{\mu} = (\psi^{\mu})^*$. In SU(3), we have 

\equb \bold{3} \otimes \bold{3} = \bold{\bar{3}} \oplus \bold{6} \eque
as indicated by
\equb \psi^{\mu}\psi^{\nu} \sim \psi^{\mu\nu} \sim  \psi^{\mu\nu}\epsilon_{\mu\nu\lambda} + \psi^{\{\mu\nu\}} \eque
Under the action of a generator $c$, $\psi^{\mu\nu}$ transforms as follows:

\equb \psi'^{\mu\nu} = c^{\mu}_{~\sigma}c^{\nu}_{~\rho}\psi^{\sigma \rho} \eque 
\\

To find the decomposition of $\bold{3_1} \otimes \bold{3_1}$, we use the generators $a_1$ and $b$ to determine how $\psi^{\{\mu\nu\}}$ decomposes.

Consider the diagonal elements $\psi^{\{11\}}, \psi^{\{22\}}, \psi^{\{33\}}$. Under $b$, they transform as

\equb \psi^{\{11\}} \rightarrow \psi^{\{22\}} \rightarrow \psi^{\{33\}} \rightarrow \psi^{\{11\}} \eque
while under $a_1$, they transform as 
\equb \psi^{\{11\}} \rightarrow \rho^2\psi^{\{11\}}, \,\,\,\,\, \psi^{\{22\}} \rightarrow \rho^6 \psi^{\{22\}}, \,\,\,\,\, \psi^{\{33\}} \rightarrow \rho^5 \psi^{\{33\}} \eque
Thus they form a $\bold{3_2}$ representation according to (\ref{a2}):

\equb \left( \begin{array}{ccc}
\psi^{\{11\}} \\
\psi^{\{22\}} \\
\psi^{\{33\}} \end{array} \right)_{\bold{3_2}} \eque
Next, consider the off-diagonal elements of the symmetric representation $\psi^{\{\mu\nu\}}$. Under $b$ they transform as

\equb \psi^{\{12\}} \rightarrow \psi^{\{23\}} \rightarrow \psi^{\{31\}} \rightarrow \psi^{\{12\}} \label{b} \eque
Under $a_1$, they transform as

\equb \psi^{\{12\}} \rightarrow \rho^4\psi^{\{12\}}, \,\,\,\,\, \psi^{\{23\}} \rightarrow \rho^{12} \psi^{\{23\}}, \,\,\,\,\, \psi^{\{31\}} \rightarrow \rho^{10} \psi^{\{31\}} \eque
thus forming the $\bold{\bar{3}_1}$ representation according to (\ref{complex}):

\equb \left( \begin{array}{ccc}
\psi^{\{23\}} \\
\psi^{\{31\}} \\
\psi^{\{12\}} \end{array} \right)_{\bold{\bar{3}_1}} \eque
We thus obtain 

\equb \bold{3_1} \otimes \bold{3_1} = \bold{\bar{3}_1} \oplus \bold{\bar{3}_1} \oplus \bold{3_2} \eque
The decompositions of $\bold{3_2} \otimes \bold{3_2}$ and $\bold{3_1} \otimes \bold{3_2}$ are found in the same way, using $a_1$ and $a_2$ appropriately. The result is displayed in Table \ref{table:Kron2}.
\\

In SU(3) we have the product between a $\bold{3}$ and a $\bold{\bar{3}}$ representation:

\equb \bold{3} \otimes \bold{\bar{3}} = \bold{1} \oplus \bold{8} \eque
where  $\psi^{\mu}\psi_{\nu} \sim \psi^{\mu}_{\nu}$ is decomposed into a trace $\psi^{\mu}_{\mu}$ that forms the singlet and a traceless ${\psi}^{\mu}_{\nu}$ that forms the adjoint 8 representation. 

For $\bold{3_1} \otimes \bold{\bar{3}_1}$, the diagonal elements of $\psi^{\mu}_{\nu}$ transform under $b$ as $\psi^1_1 \rightarrow \psi^2_2 \rightarrow \psi^3_3 \rightarrow \psi^1_1$. We have the two possibilities 

\equb \bar{\phi} = \psi^1_1+ \omega \psi^2_2 + \omega^2 \psi^3_3 \eque  
\equb \phi = \psi^1_1 + \omega^2 \psi^2_2 + \omega \psi^3_3 \eque 
Under $b$, these transform as 

\equb \bar{\phi} \rightarrow \psi^2_2 + \omega \psi^3_3 + \omega^2 \psi^1_1 = \omega^2\bar{\phi}\eque
 \equb {\phi} \rightarrow \psi^2_2 + \omega^2 \psi^3_3 + \omega \psi^1_1 = \omega{\phi} \eque 
Thus, we choose $\bar{\phi} $ to transform as $\bold{\bar{1}}'$ and $\phi $ to transform as $\bold{1'}$.

The off-diagonal terms split into two sets. They transform under $b$ in the same obvious way as described in (\ref{b}). Under $a_1$ and $a_1^*$ they transform as 
\equb \psi^1_2 \rightarrow \rho^{11} \psi^1_2, \,\,\,\,\, \psi^2_3 \rightarrow \rho^7 \psi^2_3, \,\,\,\,\, \psi^3_1 \rightarrow \rho^8 \psi^3_1 \eque
and
\equb \psi^2_1 \rightarrow \rho^{2} \psi^2_1, \,\,\,\,\, \psi^3_2 \rightarrow \rho^6 \psi^3_2, \,\,\,\,\, \psi^1_3 \rightarrow \rho^5 \psi^1_3 \eque
giving a $\bold{\bar{3}_2}$ and a $\bold{3_2}$ representation:

\equb \left( \begin{array}{ccc}
\psi^1_2 \\
\psi^2_3 \\
\psi^3_1 \end{array} \right)_{\bold{\bar{3}_2}} \eque

\equb \left( \begin{array}{ccc}
\psi^2_1 \\
\psi^3_2 \\
\psi^1_3 \end{array} \right)_{\bold{3_2}} \eque
\\

The rest of the decompositions can then be found in the same way using the appropriate representations of the generators  to investigate how the elements transform. The products between the 1D IR's are found from seeing that under $b$,  $\phi \rightarrow \omega \phi$ and $\bar{\phi} \rightarrow \omega^2 \bar{\phi}$, so that for example $\phi \bar{\phi} \rightarrow \omega \phi \omega^2 \bar{\phi} = \phi \bar{\phi}$ transforms as a $\bold{1}$. The Kronecker products are listed in table \ref{table:Kron2} and the Clebsch-Gordan decompositions in table \ref{table:Clebsch}, where we have dropped the $\{..\}$ on $\psi$.

\begin{table}[ht]
\centering      
\begin{tabular}{l}  
\hline
\hline
$\bold{1'} \otimes \bold{1'} = \bold{\bar{1}'} $ \\   
$\bold{\bar{1}'} \otimes \bold{\bar{1}'} = \bold{1'}$ \\
$\bold{1'} \otimes \bold{\bar{1}'} = \bold{1}$ \\
$\bold{3_1} \otimes \bold{3_1} = \bold{\bar{3}_1} \oplus \bold{\bar{3}_1} \oplus \bold{3_2}$   \\ 
$\bold{3_2} \otimes \bold{3_2} = \bold{\bar{3}_2} \oplus \bold{\bar{3}_1} \oplus \bold{\bar{3}_2}$   \\ 
$ \bold{3_1} \otimes \bold{\bar{3}_1} = \bold{1} \oplus \bold{1'} \oplus \bold{\bar{1}'} \oplus \bold{3_2} \oplus \bold{\bar{3}_2}$  \\
$ \bold{3_2} \otimes \bold{\bar{3}_2} = \bold{1} \oplus \bold{1'} \oplus \bold{\bar{1}'} \oplus \bold{3_1} \oplus \bold{\bar{3}_1}$  \\
$\bold{3_1} \otimes \bold{3_2} = \bold{\bar{3}_2} \oplus \bold{3_1} \oplus \bold{3_2}$   \\ 
$\bold{3_1} \otimes \bold{\bar{3}_2} = \bold{\bar{3}_2} \oplus \bold{3_1} \oplus \bold{\bar{3}_1}$   \\ 
$\bold{3_2} \otimes \bold{\bar{3}_1} = \bold{3_2} \oplus \bold{3_1} \oplus \bold{\bar{3}_1}$   \\ 
\hline     
\hline
\end{tabular} 
\caption{$Z_{13} \rtimes Z_3$ Kronecker products} 
\label{table:Kron2}  
\end{table} 

\begin{table}[ht]
\centering      
\begin{tabular}{l}  
\hline
\hline
\\
$ \bold{3_1} \otimes \bold{3_1} \rightarrow \left( \begin{array}{ccc}
\psi^{11} \\
\psi^{22} \\
\psi^{33} \end{array} \right)_{\bold{3_2}}, \,\,\,\,\, \left( \begin{array}{ccc}
\psi^{23} \\
\psi^{31} \\
\psi^{12} \end{array} \right)_{\bold{\bar{3}_1}}, \,\,\,\,\, \left( \begin{array}{ccc}
\psi^{32} \\
\psi^{13} \\
\psi^{21} \end{array} \right)_{\bold{\bar{3}_1}}$ \\   
\\
$\bold{3_2} \otimes \bold{3_2} \rightarrow \left( \begin{array}{ccc}
\psi^{22} \\
\psi^{33} \\
\psi^{11} \end{array} \right)_{\bold{\bar{3}_1}}, \,\,\,\,\, \left( \begin{array}{ccc}
\psi^{23} \\
\psi^{31} \\
\psi^{12} \end{array} \right)_{\bold{\bar{3}_2}}, \,\,\,\,\, \left( \begin{array}{ccc}
\psi^{32} \\
\psi^{13} \\
\psi^{21} \end{array} \right)_{\bold{\bar{3}_2}}$ \\
\\
$ \bold{3_1} \otimes \bold{\bar{3}_1} \rightarrow   \left( \begin{array}{ccc}
\psi^1_2 \\
\psi^2_3 \\
\psi^3_1 \end{array} \right)_{\bold{\bar{3}_2}},  \,\,\,\,\,  \left( \begin{array}{ccc}
\psi^2_1 \\
\psi^3_2 \\
\psi^1_3 \end{array} \right)_{\bold{3_2}}, \,\,\,\,\, (\psi^1_1+ \psi^2_2 + \psi^3_3)_{\bold{1}},$ \\
\\
$ \,\,\,\,\, \,\,\,\,\, \,\,\,\,\, \,\,\,\,\, \,\,\,\,\, \,\,\,\,\, (\psi^1_1+ \omega \psi^2_2 + \omega^2 \psi^3_3)_{\bold{1'}} , \,\,\,\,\,
 (\psi^1_1 + \omega^2 \psi^2_2 + \omega \psi^3_3)_{\bold{\bar{1}'}}$ \\
 \\
$ \bold{3_2} \otimes \bold{\bar{3}_2} \rightarrow   \left( \begin{array}{ccc}
\psi^3_2 \\
\psi^1_3 \\
\psi^2_1 \end{array} \right)_{\bold{\bar{3}_1}},  \,\,\,\,\,  \left( \begin{array}{ccc}
\psi^2_3 \\
\psi^3_1 \\
\psi^1_2 \end{array} \right)_{\bold{3_1}}, \,\,\,\,\, (\psi^1_1+ \psi^2_2 + \psi^3_3)_{\bold{1}},$ \\
\\ 
$  \,\,\,\,\,  \,\,\,\,\,  \,\,\,\,\,  \,\,\,\,\,  \,\,\,\,\,  \,\,\,\,\, (\psi^1_1+ \omega \psi^2_2 + \omega^2 \psi^3_3)_{\bold{1'}},  \,\,\,\,\, 
 (\psi^1_1 + \omega^2 \psi^2_2 + \omega \psi^3_3)_{\bold{\bar{1}'}}$ \\
 \\
$ \bold{3_1} \otimes \bold{3_2} \rightarrow \left( \begin{array}{ccc}
\psi^{33} \\
\psi^{11} \\
\psi^{22} \end{array} \right)_{\bold{3_1}}, \,\,\,\,\, \left( \begin{array}{ccc}
\psi^{31} \\
\psi^{12} \\
\psi^{23} \end{array} \right)_{\bold{\bar{3}_2}}, \,\,\,\,\,\left( \begin{array}{ccc}
\psi^{32} \\
\psi^{13} \\
\psi^{21} \end{array} \right)_{\bold{3_2}} $ \\
\\
$ \bold{3_1} \otimes \bold{\bar{3}_2} \rightarrow \left( \begin{array}{ccc}
\psi^1_1 \\
\psi^2_2 \\
\psi^3_3 \end{array} \right)_{\bold{\bar{3}_1}}, \,\,\,\,\,
 \left( \begin{array}{ccc}
\psi^2_3 \\
\psi^3_1 \\
\psi^1_2 \end{array} \right)_{\bold{\bar{3}_2}}, \,\,\,\,\,
 \left( \begin{array}{ccc}
\psi^2_1 \\
\psi^3_2 \\
\psi^1_3 \end{array} \right)_{\bold{3_1}}$   \\ 
\\
$\bold{3_2} \otimes \bold{\bar{3}_1} \rightarrow  \left( \begin{array}{ccc}
\psi^1_1 \\
\psi^2_2 \\
\psi^3_3 \end{array} \right)_{\bold{3_1}}, \,\,\,\,\,
 \left( \begin{array}{ccc}
\psi^1_2 \\
\psi^2_3 \\
\psi^3_1 \end{array} \right)_{\bold{\bar{3}_1}}, \,\,\,\,\,
 \left( \begin{array}{ccc}
\psi^3_2 \\
\psi^1_3 \\
\psi^2_1 \end{array} \right)_{\bold{3_2}} $   \\ 
\\
\hline     
\hline
\end{tabular} 
\caption{$Z_{13} \rtimes Z_3$ Clebsch-Gordan decompositions.} 
\label{table:Clebsch}  
\end{table} 

\section{Obtaining the mixing matrix}

We assume in this paper, that the neutrinos are Majorana type. To acquire mass, the neutrinos couple to one or more complex Higgs fields through Yukawa couplings. We have the following mass terms for the neutrinos $\nu$ and charged leptons $l$ in the Lagrangian:

\equb \mathcal{L}_{mass} = -\bar{l}_{L\alpha}(M_{l})_{\alpha \beta}l_{R\beta} - \frac{1}{2}\nu_{L\alpha}^T(M_{\nu})_{\alpha\beta}\nu_{L \beta} + h.c \eque
where $\alpha, \beta$ denote the flavour indices $e, \mu, \tau$. The masses are then obtained by finding the eigenvalues of $M_{l}$ and $M_{\nu}$ through diagonalization. 

\equb U_{L}^{\dagger}M_{l}U_R \equiv D_l =   \left(\begin{array}{ccc} m_{e} & 0 & 0  \\
0& m_{\mu}& 0\\
0 &0 & m_{\tau} \end{array} \right) \label{mixing} \eque
\equb U_{\nu}^TM_{\nu}U_{\nu} \equiv D_{\nu} =  \left(\begin{array}{ccc} m_1 & 0 & 0  \\
0& m_2& 0\\
0 &0 & m_3 \end{array} \right) \eque
The left handed charged lepton mass basis is given by:

\equb l_{L\alpha} \equiv (U_L)_{\alpha i} l_{Li} \eque
and the neutrino mass basis by:

\equb \nu_{\alpha} \equiv (U_{\nu})_{\alpha i} \nu_{i} \eque
with $i = 1,2,3$ as in (\ref{mix}), so that: 

\equb \left(\begin{array}{c}\nu_{\alpha}\\
l_{L\alpha} \end{array} \right) = \left(\begin{array}{c}U_{\nu} \nu_{i}\\
U_L l_{Li} \end{array} \right) = U_L \left(\begin{array}{c}U_L^{-1}U_{\nu} \nu_{i}\\
l_{Li} \end{array} \right) \eque
Thus, the lepton mixing matrix is given by $U = U_{L}^{\dagger}U_{\nu}$.
\\

In this paper, we adopt an effective field theory approach within a minimalist framework, assuming only SU(2) $\otimes$ U(1) as in \cite{Tetrahedral}. The charged lepton masses are generated by the dimension-4 operator :

\equb \mathcal{O}_4 = \phi^{\dagger}l^c \psi \eque
with $\phi$ the Higgs doublet, $\psi$ the lepton doublet and $l^c$ the charge conjugated lepton. Neutrino masses are generated by the dimension-5 operator \cite{Weinberg}: 

\equb \mathcal{O}_5 = (\xi_1 \tau_2 \psi) C (\xi_2 \tau_2 \psi) \label{op5} \eque
Within the Standard Model, this dimension 5 operator is the lowest dimension operator which can give masses to the neutrinos. Here the Higgs doublets $\xi_1$ and $\xi_2$ may or may not be identical and there can be as many as needed to carry out the model building \cite{Private}. We will assume that they are distinct from the Higgs doublet $\phi$. When the different (or equal) Higgs doublets $\xi_1$ and $\xi_2$ acquire a vacuum expectation value (vev) for their lower electrically neutral components, these will give masses to the neutrinos through a symmetric mass matrix.  
\\

In this paper, we do not write down the Higgs potential explicitly and minimize it. With several Higgs doublets, the potential will contain a multitude of parameters, and we will not learn much by explicitly showing how the vev's of the various Higgs fields are aligned in various regions of the parameter space. Instead, we systematically list the various possibilities and analyze the consequence for the neutrino mixing matrix. 
\\

We assume in the following, that there is no coupling between the Higgs field in the charged lepton sector and the Higgs field(s) in the neutrino sector. In other words, we have not adressed the sequestering problem \cite{Keum} \cite{Minimal}. Furthermore, we assume CP conservation, so that $M_{\nu}$ is real. 

\section{Model building}

Given the review of group theory in section 3, we are now ready to build a model for neutrino mixing based on the family symmetry group $F=Z_{13} \rtimes Z_3$. As in the standard procedure, we assign the lepton doublet fields $\psi_a$ ($a = 1,2,3$), the right handed lepton singlet fields $l_a^c$, and the Higgs fields to various representations of $T_{13}$, and write down the various allowed forms of the operators $ \mathcal{O}_4$
and $ \mathcal{O}_5$. After the Higgs fields are allowed to acquire their vev, the resulting charged lepton mass matrix $M_l$ and neutrino mass matrix $M_{\nu}$ are diagonalized, whereby the mixing matrix is obtained as outlined earlier. In order to constrain our systematic search, we will use the following assumptions during the model building:
\\
\\
i) We will align the vev of the Higgs field in the dimension-4 operator in the (1,1,1)$^T$ direction.
\\
\\
ii) We will assume no coupling between the Higgs fields in the charged lepton sector and the neutrino sector. 
\\
\\
iii) Since tribimaximal mixing gives a fairly good description of the experimental mixing matrix, at least as a first approximation, we will take it as a phenomenological guidepost. 
\\
\\
We have in mind a program in which we could eventually perturb the possible schemes, we obtain in this way, so as to study deviations from tribimaximal mixing \cite{Minimal} \cite{Harrison}.
\\

In the charged lepton sector, the operator $ \mathcal{O}_4 = \phi^{\dagger}l^c \psi$ gives a priori  $5^3 = 125$ ways of placing the three fields in different representations, but some of these are not allowed by the group theory. Also, two cases, where the representations are the conjugate of each other, will give the same result, so the number of different forms $ \mathcal{O}_4$ can acquire, is down to 27, see Table \ref{table:reps2}. For each of these cases, we then consider the various possibilities for $\mathcal{O}_5$ assigning $\xi_1$ and $\xi_2$ to various representations of $F$. We then use tribimaximal mixing as a constraint. It turns out that the cases 1, 11, 22 and 25 will lead to the same neutrino mass matrix. Another mass matrix can be found from the cases 23 and 24. Furthermore, the rest of the cases turn out to create degenerate charged lepton mass matrices, which are therefore not of phenomenological importance. Therefore, we can cover all possible results obtained from this group, by considering only four of the mentioned cases, where the lepton doublet is in two different representations. This leads to only two possible neutrino mass matrices. These cases will now be studied. In the following, we will not distinguish between upper and lower indices.
\bigskip

\begin{table}[h!]
\centering      
\begin{tabular}{c c c c}  
\hline\hline                        
Case & $\psi$ & $\phi^{\dagger}$ & $l^c$ \\  
\hline                    
1 & $\bold{3_1}$ & $\bold{3_1}$ & $\bold{3_1}$   \\   
2 &  $\bold{3_2}$ & $\bold{\bar{3}_1}$ & $\bold{\bar{3}_1}$  \\ 
3 &  $\bold{3_1}$ &  $\bold{3_1}$  & $\bold{\bar{3}_2}$  \\ 
4 &  $\bold{3_1}$ & $\bold{\bar{3}_2}$ & $\bold{3_1}$ \\  
5 & $\bold{3_2}$ & $\bold{\bar{3}_1}$&$\bold{3_1}$ \\
6 &$\bold{3_1}$  &$\bold{\bar{3}_1}$ & $\bold{3_2}$ \\ 
7 &$\bold{3_2}$ &$\bold{3_1}$ & $\bold{\bar{3}_1}$\\
8 & $\bold{3_1} $ & $\bold{\bar{3}_1} $ & $\bold{\bar{3}_2}$ \\
9 &$\bold{3_1}$ & $\bold{3_2}$ & $\bold{\bar{3}_1}$ \\
10& $\bold{3_1}$& $\bold{\bar{3}_2}$& $\bold{\bar{3}_1}$\\
11&$\bold{3_2}$ &$\bold{3_2}$ &$\bold{3_2}$ \\
12&$\bold{3_1}$ &$\bold{3_2}$ & $\bold{3_2}$\\
13&$\bold{3_2}$ &$\bold{3_2}$ & $\bold{3_1}$\\
14&$\bold{3_2}$ &$\bold{3_1}$ &$\bold{3_2}$ \\
15&$\bold{3_1}$ &$\bold{3_2}$ & $\bold{\bar{3}_2}$\\
16 &$\bold{3_2}$ &$\bold{\bar{3}_2}$ &$\bold{\bar{3}_1}$ \\
17&$\bold{3_2}$ &$\bold{\bar{3}_1}$ & $\bold{\bar{3}_2}$\\
18&$\bold{3_2}$ & $\bold{3_1}$& $\bold{\bar{3}_2}$\\
19& $\bold{3_2}$&$\bold{\bar{3}_2}$ & $\bold{3_1}$  \\
20&$\bold{3_1}$ & $\bold{\bar{3}_2}$&$\bold{3_2}$ \\ 
21&$\bold{1}, \bold{1'}, \bold{\bar{1}'}$ &$\bold{3_1}$ & $\bold{\bar{3}_1}$\\
22&$\bold{3_1}$ & $\bold{\bar{3}_1}$& $\bold{1}, \bold{1'}, \bold{\bar{1}'}$ \\
23&$\bold{3_1}$ & $\bold{1}, \bold{1'}, \bold{\bar{1}'} $& $\bold{\bar{3}_1}$\\
24& $\bold{3_2}$& $\bold{1}, \bold{1'}, \bold{\bar{1}'}$ & $\bold{\bar{3}_2}$\\
25&$\bold{3_2}$ &$\bold{\bar{3}_2}$ & $\bold{1}, \bold{1'}, \bold{\bar{1}'}$ \\
26&$\bold{1}, \bold{1'}, \bold{\bar{1}'}$  &$\bold{3_2}$ & $\bold{\bar{3}_2}$\\
27 &$ \bold{1}, \bold{1'}, \bold{\bar{1}'}$ &$\bold{1}, \bold{1'}, \bold{\bar{1}'} $ &$\bold{1}, \bold{1'}, \bold{\bar{1}'}$ \\

\hline     
\end{tabular} 
\caption{Possible representations for the 3 fields.} 
\label{table:reps2}  
\end{table} 

\section{$\bold{Case\, 1}$}

The charge conjugated lepton, $l^c$, and the lepton doublet, $\psi$, both in $\bold{3_1}$ can be combined in two ways, 

\equb \nonumber \bold{\bar{3}_1}: \,\,\,\,\, \{l^c_2\psi_3, l^c_3\psi_1, l^c_1\psi_2 \} \eque
\equb \bold{\bar{3}_1}: \,\,\,\,\, \{l^c_3\psi_2, l^c_1\psi_3, l^c_2\psi_1\} \eque
With $\phi^{\dagger}$ in $\bold{3_1}$, we get the invariant terms in the dimension 4 operator of the Lagrangian:

\equb h_1(\phi^{\dagger}_1l^c_2\psi_3 + \phi^{\dagger}_2 l^c_3\psi_1 + \phi^{\dagger}_3 l^c_1\psi_2) + h_2(\phi^{\dagger}_1l^c_3\psi_2 + \phi^{\dagger}_2 l^c_1\psi_3 + \phi^{\dagger}_3 l^c_2\psi_1) \eque 
\equb \nonumber= (l_1^c,l_2^c,l_3^c) \left( \begin{array}{ccc}
0 & h_1\phi^{\dagger}_3& h_2\phi^{\dagger}_2 \\
h_2 \phi^{\dagger}_3 & 0 & h_1\phi^{\dagger}_1\\
h_1 \phi^{\dagger}_2 & h_2 \phi^{\dagger}_1 & 0 \end{array} \right)  \left( \begin{array}{ccc}
\psi_1\\
\psi_2\\
\psi_3 \end{array} \right) \eque
Upon spontaneous symmetry breaking, we let the Higgs field acquire a vev in the $(1,1,1)^T$ direction, and the mass matrix becomes:

\equb M_l = v\left( \begin{array}{ccc}
0 & h_1& h_2 \\
h_2 &0& h_1\\
h_1 & h_2 & 0 \end{array} \right) \eque
To diagonalize $M_l$, we multiply by its hermitian conjugate:

\equb M_lM_l^{\dagger} = M_l^{\dagger}M_l = \left( \begin{array}{ccc}
\mid h_1 \mid^2+ \mid h_2 \mid^2 & h_1^*h_2& h_1h_2^* \\
h_1h_2^* &\mid h_1 \mid^2 + \mid h_2 \mid^2& h_1^*h_2\\
h_1^*h_2 & h_1h_2^* & \mid h_1 \mid^2 + \mid h_2 \mid^2 \end{array} \right) \eque
This is a circulant matrix, that can be diagonalized by the following unitary matrix:

\equb U_L^{\dagger} = \frac{1}{\sqrt{3}}\left( \begin{array}{ccc}
1 & 1& 1 \\
\omega &1& \omega^2\\
\omega^2 & 1 & \omega \end{array} \right) \eque

We now look at the dimension-5 operator and ask what Higgs field vev's will lead to tribimaximal mixing.\footnote{In order to find the different invariant terms that enter the Lagrangian through this operator, it is necessary to keep in mind, that since the neutrinos are Majorana, we have here used the 2 component spinor notation, so that each lepton and scalar carries the Lorentz spinor indices $\alpha = 1,2$. For example, $\psi\psi = \psi^{\alpha}\psi_{\alpha} = \epsilon_{\alpha\beta}\psi^{\alpha}\psi^{\beta}$ \cite{Bentov}. We also have to take into account that the leptons are Grassman numbers, and thereby anticommute.} The Higgs fields $\xi_1$ and $\xi_2$ in (\ref{op5}) can each be any of the following 5 different Higgs in the representations: 

\equb \chi \sim \bold{1}, \,\,\,\,\, \eta \sim \bold{3_1}, \,\,\,\,\, \rho \sim \bold{3_2}, \,\,\,\,\, \kappa \sim \bold{\bar{3}_1}, \,\,\,\,\, \zeta \sim \bold{\bar{3}_2} \eque
We have a priori $5+4+3+2+1=15$ combinations of these in the dimension-5 operator. In Case 1, we have the $\psi$ in the $\bold{3_1}$ representation. From Table \ref{table:Kron2}, we can already exclude the combinations $(\chi \psi)^2$, $(\eta \psi)^2$, $(\chi \psi)(\kappa \psi)$ and $(\chi \psi)(\rho \psi)$. The possible invariants from combining two lepton doublets with two Higgs in the above mentioned representations and contracting the indices properly,  can now be cataloged. 
\\
\\
a) $(\rho \psi)^2$: 
\equb f_1 (\rho_3 \psi_2 \rho_1 \psi_2  + \rho_1 \psi_3 \rho_2 \psi_3  + \rho_2 \psi_1 \rho_3 \psi_1 ) \eque
b) $(\kappa \psi)^2$: 
\equb  \nonumber f_2(\kappa_1\psi_1\kappa_1\psi_1 + \kappa_2\psi_2\kappa_2\psi_2 + \kappa_3\psi_3\kappa_3\psi_3) \eque
\equb \nonumber + f_3(\kappa_1\psi_1\kappa_2\psi_2 + \kappa_3\psi_3\kappa_1\psi_1 + \kappa_2\psi_2\kappa_3\psi_3) \eque
\equb + f_4(\kappa_1\psi_2\kappa_2\psi_1 + \kappa_2\psi_3\kappa_3\psi_2 + \kappa_3\psi_1\kappa_1\psi_3) \eque
c) $(\zeta \psi)^2$:
\equb f_5(  \zeta_1\psi_1\zeta_1\psi_2 + \zeta_2\psi_2\zeta_2\psi_3 + \zeta_3\psi_3\zeta_3\psi_1) \eque
d) $(\chi \psi )( \eta \psi)$:   
\equb \nonumber f_6 \chi(\psi_1 \eta_2 \psi_3 + \psi_2 \eta_3 \psi_1 + \psi_3 \eta_1 \psi_2) \eque
\equb+ f_7 \chi(\psi_1 \eta_3 \psi_2 + \psi_2 \eta_1 \psi_3 + \psi_3 \eta_2 \psi_1) \eque
e) $(\chi \psi)(\zeta \psi)$:
\equb f_8 \chi(\psi_1 \zeta_1 \psi_1 + \psi_2 \zeta_2 \psi_2 + \psi_3 \zeta_3 \psi_3) \eque
f) $(\eta \psi)( \rho \psi)$:
\equb \nonumber f_9 (\eta_2 \psi_3 \rho_3 \psi_3 + \eta_3 \psi_1 \rho_1 \psi_1 + \eta_1 \psi_2 \rho_2 \psi_2) \eque
\equb \nonumber f_{10} (\eta_3 \psi_2 \rho_3 \psi_3 + \eta_1 \psi_3 \rho_1 \psi_1 + \eta_2 \psi_1\rho_2 \psi_2) \eque
\equb + f_{11}(\eta_3 \psi_3 \rho_3 \psi_2 + \eta_1 \psi_1 \rho_1 \psi_3 + \eta_2 \psi_2 \rho_2 \psi_1) \eque
g) $(\eta \psi)(\kappa \psi)$:
\equb f_{12} (\eta_1 \psi_1 \kappa_2 \psi_1 + \eta_2 \psi_2 \kappa_3 \psi_2 + \eta_3 \psi_3 \kappa_1 \psi_3) \eque
h) $(\eta \psi)(\zeta \psi)$:
\equb  \nonumber f_{13} (\eta_3\psi_2\zeta_1\psi_2 + \eta_1\psi_3\zeta_2\psi_3 + \eta_2\psi_1\zeta_3\psi_1 ) \eque
\equb \nonumber f_{14} (\eta_2\psi_3\zeta_1\psi_2 + \eta_3\psi_1\zeta_2\psi_3 + \eta_1\psi_2\zeta_3\psi_1) \eque 
\equb + f_{15}(\eta_1\psi_1\zeta_3\psi_2 + \eta_2\psi_2\zeta_1\psi_3 + \eta_3\psi_3\zeta_2\psi_1) \eque
i) $(\rho \psi)(\kappa \psi) $:
\equb \nonumber f_{16} (\rho_2\psi_3\kappa_2\psi_1 + \rho_3\psi_1\kappa_3\psi_2 + \rho_1\psi_2\kappa_1\psi_3) \eque
\equb + f_{17}(\rho_1\psi_3\kappa_1\psi_2 + \rho_2\psi_1\kappa_2\psi_3 + \rho_3\psi_2\kappa_3\psi_1) \eque
j) $(\rho \psi)(\zeta \psi)$:
\equb \nonumber f_{18} (\rho_2\psi_3\zeta_3\psi_2 + \rho_3\psi_1\zeta_1\psi_3 + \rho_1\psi_2\zeta_2\psi_1)\eque
\equb  + f_{19}(\rho_3\psi_3\zeta_1\psi_1 + \rho_1\psi_1\zeta_2\psi_2 + \rho_2\psi_2\zeta_3\psi_3) \eque
k) $(\kappa\psi)(\zeta \psi)$: 
\equb f_{20} (\kappa_1\psi_2\zeta_3\psi_2 + \kappa_2\psi_3\zeta_1\psi_3 + \kappa_3\psi_1\zeta_2\psi_1) \eque
Of course, in building a specific model, we are free to omit one or more of these Higgs fields, narrowing down the possibilities further. There should to begin with be 5 different terms in b), since $\bold{3_1} \otimes \bold{\bar{3}_1} = \bold{1} \oplus \bold{1'} \oplus \bold{\bar{1}'} \oplus \bold{3_2} \oplus \bold{\bar{3}_2}$ multiplied by itself creates invariants from $\bold{1} \otimes \bold{1}$, $\bold{1'} \otimes \bold{\bar{1}'}$, $\bold{3_2} \otimes \bold{\bar{3}_2}$ and two charge conjugates. However, when multiplying out the terms and seeing that $\omega + \omega^2 = -1$, we get the three listed terms. 
\\

In Case 1, we first consider including only the Higgs fields $\eta$ and $\kappa$, to which we assign the vev's:
\\

i)
 \equb \langle \eta \rangle = v_1 \left( \begin{array}{ccc}
0 \\ 1  \\ 0 \end{array} \right), \,\,\,\,\,  \langle \kappa \rangle = u_1\left( \begin{array}{ccc}
1 \\ 0  \\ 1 \end{array} \right) \label{vev} \eque 
so that we get contributions from invariants in b) and g). We could also add $\chi$ so we have the vev's:
\\

ii)
 \equb \langle \chi \rangle = x, \,\,\,\,\, \langle \eta \rangle = v_1 \left( \begin{array}{ccc}
0 \\ 1  \\ 0 \end{array} \right), \,\,\,\,\,  \langle \kappa \rangle = u_1\left( \begin{array}{ccc}
1 \\ 0  \\ 1 \end{array} \right) \eque 
(with $\eta$ and $\kappa$ having the same vev's as in (\ref{vev})). This would give another contribution from d).

Upon spontaneous symmetry breaking, the neutral lower components of the SU(2) Higgs doublets can acquire these vev's and so will connect with the neutrinos giving masses to these. The charge conjugation operator, together with the neutrinos being Grassman, will then allow us to combine some of the coupling constants in b), d), f), h), i) and j), contributing to the off diagonal elements of the neutrino mass matrix, thus making this symmetric as it should be. In the following equations, we will, in some cases, combine $f_a$ and $f_{a+1}$ into a new coupling constant, which we call $f'_a$. The terms contributing to the neutrino mass matrix in i) are then given by:

\equb  f_2u_1^2(\nu_1\nu_1 + \nu_3\nu_3) + f'_3u_1^2\nu_1\nu_3 + f_{12}v_1u_1\nu_2\nu_2 \eque
and for $\langle \chi \rangle$ getting a vev in ii), we also get the contribution:

\equb f'_6xv_1\nu_1\nu_3 \eque
This leads to the possible mass matrices:
\\

i)
\equb M_{\nu} =  \left( \begin{array}{ccc}
f_2u_1^2 & 0 & f'_3u_1^2  \\
0 & f_{12}v_1u_1 & 0 \\
f'_3u_1^2 & 0 & f_2u_1^2 \end{array} 
\right) \eque
and
\\

ii)
\equb M_{\nu} =  \left( \begin{array}{ccc}
f_2u_1^2 & 0 & f'_3u_1^2 + f'_6xv_1\\
0 & f_{12}v_1u_1 & 0 \\
f'_3u_1^2 + f'_6xv_1& 0 & f_2u_1^2 \end{array} 
\right) \eque
with and without the Higgs $\chi$ respectively, both having the form:

\equb M_{\nu} = \left( \begin{array}{ccc}
\alpha & 0 & \beta \\
0 &\gamma& 0 \\
\beta & 0 & \alpha \end{array} \right) \label{mass} \eque
which is diagonalized by the unitary matrix:

\equb U_{\nu} = \frac{1}{\sqrt{2}} \left( \begin{array}{ccc}
1 & 0& -1 \\
0 &\sqrt{2}& 0\\
1 & 0 & 1 \end{array} \right) \eque
We can now find the lepton mixing matrix:

\equb \nonumber U = U_L^{\dagger}U_{\nu} = \frac{1}{\sqrt{3}}\left( \begin{array}{ccc}
1 & 1& 1 \\
\omega &1& \omega^2\\
\omega^2 & 1 & \omega \end{array} \right) \frac{1}{\sqrt{2}} \left( \begin{array}{ccc}
1 & 0& -1 \\
0 &\sqrt{2}& 0\\
1 & 0 & 1 \end{array} \right)  \eque
\equb = \left( \begin{array}{ccc}
\frac{2}{\sqrt{6}} & \frac{1}{\sqrt{3}} & 0 \\
-\frac{1}{\sqrt{6}} & \frac{1}{\sqrt{3}} & \frac{1}{\sqrt{2}} \\
-\frac{1}{\sqrt{6}} & \frac{1}{\sqrt{3}} & -\frac{1}{\sqrt{2}} \end{array} \right) = U_{TB} \eque
This is an interesting result. The neutrino mass matrix in (\ref{mass}), which leads to tribimaximal mixing in case 1, is exactly achieved, by choosing the right vev's in the neutrino sector. This was not possible in \cite{Tetrahedral} for the tetrahedral group. In this paper, when choosing vev's for the Higgs fields in the neutrino sector, the exact form of the mass matrix in (\ref{mass}) could not be achieved. The mass matrix had to be perturbed in the following way:

\equb M_{\nu} = \left( \begin{array}{ccc}
\alpha +\epsilon& 0 & \beta \\
0 &\gamma& 0 \\
\beta & 0 & \alpha - \epsilon \end{array} \right) \eque
which did not allow for exact tribimaximal mixing. It was instead assumed, that $\epsilon$ being small, could be seen as a perturbation, and thereby only lead to a small deviation from tribimaximal. If tribimaximal mixing is the best fit for lepton mixing, we already here have evidence, that the Frobenius group can be used to describe neutrino mixing. In order to determine which of the two groups, the tetrahedral or the Frobenius, is the best candidate to describe neutrino mixing, it will be necessary to test the Frobenius group further, using more constraints. 

\section{Case 23:}

The lepton doublets are in the $\bold{3_1}$ representation, the charged lepton singlets in the $\bold{\bar{3}_1}$ representation, and the Higgs bosons $\phi^{\dagger}$ in the $\bold{1}$, $\bold{1'}$ and $\bar{\bold{1}'}$ representations. $\bold{3_1}$ and $\bold{\bar{3}_1}$ can be combined to form the 3 1 dimensional representations:

\equb \nonumber \bold{1} \simeq \psi_1l^c_1 + \psi_2l^c_2 + \psi_3l^c_3 \eque
\equb \nonumber  \bold{1}' \simeq \psi_1l^c_1 + \omega \psi_2l^c_2 + \omega^2\psi_3l^c_3 \eque
\equb  \bold{\bar{1}}' \simeq \psi_1l^c_1 + \omega^2\psi_2l^c_2 +\omega \psi_3l^c_3 \eque
Combined with the $\phi^{\dagger}$, we get the following contributions:

\equb \nonumber (l_1^c,l_2^c,l_3^c) \left( \begin{array}{ccc}
h_1\phi^{\dagger}_1 + h_2\omega\phi^{\dagger}_2 + h_3\omega^2\phi^{\dagger}_3 & 0 & 0 \\
0 & h_1\phi^{\dagger}_1 + h_2\phi^{\dagger}_2 + h_3\phi^{\dagger}_3 & 0 \\
0 & 0 & h_1\phi^{\dagger}_1 + h_2\omega^2\phi^{\dagger}_2 + h_3\omega\phi^{\dagger}_3 \end{array} \right) \eque
\equb \times \left( \begin{array}{ccc}
\psi_1\\
\psi_2\\
\psi_3 \end{array} \right) \eque
When the Higgs field $\phi^{\dagger}$ again acquires a vev in the $(1,1,1)^T$ direction, the charged lepton mass matrix is already diagonal. To get tribimaximal lepton mixing, we then need the neutrino mixing matrix to be tribimaximal. We use the invariants as listed for case 1 above for the $\psi$ in the $\bold{3_1}$ representation to achieve this by letting the Higgs bosons acquire the following vev's:
\\

i)
\equb \nonumber \langle \eta \rangle = v_1 \left( \begin{array}{ccc}
0 \\ 1  \\ 1 \end{array} \right), \,\,\,\,\, \langle \rho \rangle = v_2 \left( \begin{array}{ccc}
1 \\ 0  \\ 0 \end{array} \right) \eque
\equb  \langle \kappa \rangle = u_1\left( \begin{array}{ccc}
1 \\ 1 \\ 1 \end{array} \right)\eque 
This will give contributions from b), f), g) and i). Letting $\chi$ acquire a vev in addition to these could also be a possibility:
\\

ii)
\equb \nonumber \langle \chi \rangle = x, \,\,\,\,\,\langle \eta \rangle = v_1 \left( \begin{array}{ccc}
0 \\ 1  \\ 1 \end{array} \right), \,\,\,\,\, \langle \rho \rangle = v_2 \left( \begin{array}{ccc}
1 \\ 0  \\ 0 \end{array} \right) \eque
\equb  \langle \kappa \rangle = u_1\left( \begin{array}{ccc}
1 \\ 1 \\ 1 \end{array} \right)\eque 
thereby gaining a contribution from d). Writing up these terms in the neutrino mass matrix, we find:
\\

i)
\equb M_{\nu} =  \left( \begin{array}{ccc}
f_2u_1^2 + f_9v_1v_2 & f'_3u_1^2 & f'_3u_1^2  \\
f'_3u_1^2 & f_2u_1^2+ f_{12}v_1u_1 & f'_3u_1^2 + f'_{16}v_2u_1 \\
f'_3u_1^2 & f'_3u_1^2 + f'_{16}v_2u_1 & f_2u_1^2 + f_{12}v_1u_1 \end{array} 
\right) \eque
and
\\

ii)
\equb M_{\nu} =  \left( \begin{array}{ccc}
f_2u_1^2 + f_9v_1v_2 & f'_3u_1^2 +f'_6xv_1& f'_3u_1^2 +f'_6xv_1 \\
f'_3u_1^2 +f'_6xv_1 & f_2u_1^2+ f_{12}v_1u_1 & f'_3u_1^2 + f'_{16}v_2u_1 \\
f'_3u_1^2 + f'_6xv_1 & f'_3u_1^2 + f'_{16}v_2u_1 & f_2u_1^2 + f_{12}v_1u_1 \end{array} 
\right) \eque
leading to a matrix with $\mu-\tau$ symmetry:

\equb M_{\nu} =  \left( \begin{array}{ccc}
\gamma & \beta & \beta  \\
\beta & \alpha & \delta \\
\beta & \delta & \alpha \end{array} 
\right) \eque
which is diagonalized by the tribimaximal mixing matrix, if we impose without justification, $\delta = \gamma + \beta - \alpha$.

\section{Case 25:}

We now have the $l^c$ in the three 1D representations, the lepton doublet in the $\bold{3_2}$ representation and the $\phi^{\dagger}$ in the $\bold{\bar{3}_2}$ representation. The lepton doublet and the Higgs then join to form the three 1 dimensional representations:

\equb \nonumber \bold{1} \simeq \phi_1^{\dagger}\psi_1 + \phi_2^{\dagger}\psi_2 + \phi_3^{\dagger}\psi_3 \eque
\equb \nonumber  \bold{1}' \simeq \phi_1^{\dagger}\psi_1 + \omega \phi_2^{\dagger}\psi_2 + \omega^2\phi_3^{\dagger}\psi_3 \eque
\equb  \bold{\bar{1}}' \simeq \phi_1^{\dagger}\psi_1 + \omega^2\phi_2^{\dagger}\psi_2 +\omega \phi_3^{\dagger}\psi_3 \eque
which results in invariants when connected with $\bold{1}\simeq l_1^c, \bold{\bar{1}}' \simeq l_3^c $ and $\bold{1}' \simeq  l_2^c$ respectively. The mass terms of the charged leptons in the Lagrangian are then (using the freedom to incorporate phases into the coupling constants $h_i$):

\equb \nonumber h_1l_1^c(\phi_1^{\dagger}\psi_1 + \phi_2^{\dagger}\psi_2 + \phi_3^{\dagger}\psi_3) + h_3\omega^2l_3^c( \phi_1^{\dagger}\psi_1 + \omega \phi_2^{\dagger}\psi_2 + \omega^2\phi_3^{\dagger}\psi_3) \eque
\equb \nonumber+ h_2\omega l_2^c(\phi_1^{\dagger}\psi_1 + \omega^2\phi_2^{\dagger}\psi_2 +\omega \phi_3^{\dagger}\psi_3) = h_1l_1^c(\phi_1^{\dagger}\psi_1 + \phi_2^{\dagger}\psi_2 + \phi_3^{\dagger}\psi_3) \eque
\equb \nonumber + h_2 l_2^c(\omega \phi_1^{\dagger}\psi_1 + \phi_2^{\dagger}\psi_2 +\omega^2 \phi_3^{\dagger}\psi_3)+ h_3l_3^c(\omega^2 \phi_1^{\dagger}\psi_1 + \phi_2^{\dagger}\psi_2 + \omega\phi_3^{\dagger}\psi_3) \eque
\equb \nonumber = \left( \begin{array}{ccc}
l_1^c & l_2^c& l_3^c  \end{array} \right) \left( \begin{array}{ccc}
h_1 & 0& 0 \\
0 &h_2& 0 \\
0 & 0 & h_3 \end{array} \right) \left( \begin{array}{ccc}
1 & 1& 1 \\
\omega &1& \omega^2\\
\omega^2 & 1 & \omega \end{array} \right)\eque 
\equb \times \left( \begin{array}{ccc}
\phi^{\dagger}_1 &0& 0 \\
0 &\phi^{\dagger}_2& 0\\
0 & 0 & \phi^{\dagger}_3 \end{array} \right) \left( \begin{array}{ccc}
\psi_1 \\
\psi_2\\
\psi_3 \end{array} \right) \eque
Upon spontaneous symmetry breaking, again with the Higgs vev in the $(1,1,1)^T$ direction, the mass matrix is diagonalized by the same unitary matrix as in case 1:

\equb U_L^{\dagger} =  \frac{1}{\sqrt{3}}\left( \begin{array}{ccc}
1 & 1& 1 \\
\omega &1& \omega^2\\
\omega^2 & 1 & \omega \end{array} \right) \eque

Proceeding to the neutrino sector, we now have $\psi$ in the $\bold{3_2}$ representation, leading to different invariants in the dimension-5 operator than for case 1 and 3.  The Higgs fields $\xi_1$ and $\xi_2$ in (\ref{op5}) can each still be any of the following 5 Higgs with the assigned representations: 

\equb \chi \sim \bold{1}, \,\,\,\,\, \eta \sim \bold{3_1}, \,\,\,\,\, \rho \sim \bold{3_2}, \,\,\,\,\, \kappa \sim \bold{\bar{3}_1}, \,\,\,\,\, \zeta \sim \bold{\bar{3}_2} \eque
Using table \ref{table:Kron2} once again, we can omit the combinations $(\chi\psi)^2$, $(\rho\psi)^2$, $(\chi\psi)(\zeta\psi)$ and $(\chi\psi)(\kappa\psi)$. This leaves the possibilities:
\\
\\
a)  $(\eta \psi)^2$: 
\equb f_1 ( \eta_3 \psi_1 \eta_3 \psi_2  + \eta_1 \psi_2 \eta_1 \psi_3  + \eta_2 \psi_3 \eta_2 \psi_1 ) \eque
b) $(\kappa \psi)^2$: 
\equb f_2(\kappa_1\psi_1\kappa_2\psi_1 + \kappa_2\psi_2\kappa_3\psi_2 + \kappa_3\psi_3\kappa_1\psi_3) \eque
c) $(\zeta \psi)^2$:
\equb \nonumber   f_3(\zeta_1\psi_1\zeta_1\psi_1 + \zeta_2\psi_2\zeta_2\psi_2 + \zeta_3\psi_3\zeta_3\psi_3) \eque
\equb  \nonumber + f_4(\zeta_1\psi_1\zeta_2\psi_2 +  \zeta_3\psi_3\zeta_1\psi_1 + \zeta_2\psi_2\zeta_3\psi_3) \eque
\equb + f_5(\zeta_1\psi_2\zeta_2\psi_1 +  \zeta_3\psi_1\zeta_1\psi_3 + \zeta_2\psi_3\zeta_3\psi_2) \eque
d) $(\chi \psi )( \eta \psi)$:   
\equb f_6 \chi  (\psi_1 \eta_3 \psi_1 + \psi_2 \eta_1 \psi_2 + \psi_3 \eta_2 \psi_3) \eque
e) $(\chi \psi)(\rho \psi)$: 
\equb \nonumber f_7 \chi(\psi_1 \rho_2 \psi_3 + \psi_2 \rho_3 \psi_1 + \psi_3 \rho_1 \psi_2) \label{combinef6}\eque
\equb + f_8\chi(\psi_1 \rho_3 \psi_2 + \psi_2 \rho_1 \psi_3 + \psi_3 \rho_2 \psi_1) \label{combinef7} \eque
f) $(\eta \psi)( \rho \psi)$:
\equb \nonumber  f_9 (\eta_3 \psi_2 \rho_3 \psi_2 + \eta_1 \psi_3 \rho_1 \psi_3 + \eta_2 \psi_1 \rho_2 \psi_1) \eque
\equb \nonumber f_{10} (\eta_3 \psi_2 \rho_2 \psi_3 + \eta_1 \psi_3 \rho_3 \psi_1 + \eta_2 \psi_1\rho_1 \psi_2) \label{combinef9}\eque 
\equb + f_{11}(\eta_3 \psi_3 \rho_2 \psi_2 + \eta_1 \psi_1 \rho_3 \psi_3 + \eta_2 \psi_2 \rho_1 \psi_1) \label{combinef10} \eque
g) $(\eta \psi)(\kappa \psi)$:
\equb \nonumber f_{12} (\eta_3 \psi_1 \kappa_2 \psi_3 + \eta_1\psi_2 \kappa_3 \psi_1 + \eta_2 \psi_3 \kappa_1 \psi_2)\label{combinef11}\eque
\equb + f_{13}(\eta_3 \psi_3 \kappa_2 \psi_1 + \eta_1 \psi_1 \kappa_3 \psi_2 + \eta_2 \psi_2 \kappa_1 \psi_3)\label{combinef12} \eque
h) $(\eta \psi)(\zeta \psi)$:
\equb  f_{14} (\eta_3\psi_3\zeta_2\psi_3 + \eta_1\psi_1\zeta_3\psi_1 + \eta_2\psi_2\zeta_1\psi_2) \eque
i) $(\rho \psi)(\kappa \psi) $:
\equb \nonumber f_{15} (\rho_3\psi_2\kappa_2\psi_3 + \rho_1\psi_3\kappa_3\psi_1 + \rho_2\psi_1\kappa_1\psi_2) \label{combinef14}\eque
\equb \nonumber + f_{16}(\rho_2\psi_2\kappa_1\psi_1 + \rho_3\psi_3\kappa_2\psi_2 + \rho_1\psi_1\kappa_3\psi_3) \label{combinef15} \eque
\equb f_{17} (\rho_2\psi_3\kappa_2\psi_3 + \rho_3\psi_1\kappa_3\psi_1 + \rho_1\psi_2\kappa_1\psi_2) \eque
j) $(\rho \psi)(\zeta \psi)$:
\equb f_{18} (\rho_2\psi_2\zeta_3\psi_2 + \rho_3\psi_3\zeta_1\psi_3 + \rho_1\psi_1\zeta_2\psi_1) \eque
k) $(\kappa\psi)(\zeta \psi)$: 
\equb \nonumber f_{19} (\kappa_1\psi_1\zeta_2\psi_3 + \kappa_2\psi_2\zeta_3\psi_1 + \kappa_3\psi_3\zeta_1\psi_2) \label{combinef18}\eque
\equb + f_{20}(\kappa_2\psi_1\zeta_3\psi_2 + \kappa_3\psi_2\zeta_1\psi_3 + \kappa_1\psi_3\zeta_2\psi_1) \label{combinef19}\eque
Evidently, the coefficients here are not the same as for $\psi$ in the $\bold{3_1}$ representation. In this case, the possible Higgs field vev's needed to eventually get tribimaximal mixing are numerous. We list them all in the following:
\\

i)
 \equb \langle \rho \rangle = v_2 \left( \begin{array}{ccc}
0 \\ 1  \\ 0 \end{array} \right), \,\,\,\,\, \langle \zeta \rangle = u_2\left( \begin{array}{ccc}
1 \\ 0  \\ 1 \end{array} \right) \eque 
leading to contributions from c) and j).

ii)
  \equb \langle \chi \rangle = x, \,\,\,\,\, \langle \rho \rangle = v_2 \left( \begin{array}{ccc}
0 \\ 1  \\ 0 \end{array} \right), \,\,\,\,\, \langle \zeta \rangle = u_2\left( \begin{array}{ccc}
1 \\ 0  \\ 1 \end{array} \right) \eque 
which gives the same contributions as in i) with one extra from e).

iii)
\equb \langle \kappa \rangle = u_1\left( \begin{array}{ccc}
1 \\ 1  \\ 1 \end{array} \right), \,\,\,\,\, \langle \zeta \rangle = u_2\left( \begin{array}{ccc}
0 \\ 1  \\ 0 \end{array} \right) \eque 
with b), c) and k) contributing to the mass matrix.

iv)
\equb \langle \chi \rangle = x, \,\,\,\,\, \langle \kappa \rangle = u_1\left( \begin{array}{ccc}
1 \\ 1  \\ 1 \end{array} \right), \,\,\,\,\, \langle \zeta \rangle = u_2\left( \begin{array}{ccc}
0 \\ 1  \\ 0 \end{array} \right) \eque 
leading to the same result as in iii) with no extra contribution. 

v) 
\equb  \langle \rho \rangle = v_2 \left( \begin{array}{ccc}
1 \\ 0  \\ 0 \end{array} \right), \,\,\,\,\, \langle \kappa \rangle = u_1\left( \begin{array}{ccc}
1 \\ 1  \\ 1 \end{array} \right) \eque 
giving terms to the mass matrix from b) and i).

vi) 
\equb \langle \eta \rangle = v_1 \left( \begin{array}{ccc}
0 \\ 1  \\ 0 \end{array} \right), \,\,\,\,\, \langle \zeta \rangle = u_2\left( \begin{array}{ccc}
1 \\ 0  \\ 1 \end{array} \right) \eque 
contributing with terms from a), c) and h). 
\\

When the neutral component of the Higgs bosons acquire these vev's we can then combine the coupling constants as we did for the invariants with $\psi \sim \bold{3_1}$. The possible mass matrices from these couplings and vevs will then become: 
\\

i)
\equb M_{\nu} =  \left( \begin{array}{ccc}
f_3u_2^2  & 0 & f'_4u_2^2  \\
0 & f_{18}v_2u_2 & 0 \\
f'_4u_2^2 & 0 & f_3u_2^2 \end{array} 
\right) \eque

ii)
\equb M_{\nu} =  \left( \begin{array}{ccc}
f_3u_2^2  & 0 & f'_4u_2^2 +f'_7xv_2 \\
0 & f_{18}v_2u_2 & 0 \\
f'_4u_2^2 +f'_7xv_2 & 0 & f_3u_2^2 \end{array} 
\right) \eque

iii), iv)
\equb M_{\nu} =  \left( \begin{array}{ccc}
f_2u_1^2  & 0 & f'_{19}u_1u_2  \\
0 & f_{2}u_1^2 + f_3u_2^2 & 0 \\
f'_{19}u_1u_2 & 0 & f_2u_1^2 \end{array} 
\right) \eque

v)
\equb M_{\nu} =  \left( \begin{array}{ccc}
f_2u_1^2  & 0 & f'_{15}v_2u_1  \\
0 & f_{2}u_1^2 + f_{17}v_2u_1 & 0 \\
f'_{15}v_2u_1 & 0 & f_2u_1^2 \end{array} 
\right) \eque

vi)
\equb M_{\nu} =  \left( \begin{array}{ccc}
f_3u_2^2  & 0 & f_1v_1^2 + f'_4u_2^2  \\
0 & f_{14}v_1u_2 & 0 \\
f_1v_1^2 + f'_4u_2^2 & 0 & f_3u_2^2 \end{array} 
\right) \eque
all with the same form as in case 1:

\equb M_{\nu} = \left( \begin{array}{ccc}
\alpha & 0& \beta \\
0 &\gamma& 0\\
\beta & 0 & \alpha \end{array} \right) \eque
and as in case 1, tribimaximal mixing is possible from this mass matrix. 

\section{Case 24:}

This case gives the same result as case 23 in the charged lepton sector. We therefore have the following contributions to the charged lepton mass matrix:

\equb \nonumber (l_1^c,l_2^c,l_3^c) \left( \begin{array}{ccc}
h_1\phi^{\dagger}_1 + h_2\omega\phi^{\dagger}_2 + h_3\omega^2\phi^{\dagger}_3 & 0 & 0 \\
0 & h_1\phi^{\dagger}_1 + h_2\phi^{\dagger}_2 + h_3\phi^{\dagger}_3 & 0 \\
0 & 0 & h_1\phi^{\dagger}_1 + h_2\omega^2\phi^{\dagger}_2 + h_3\omega\phi^{\dagger}_3 \end{array} \right)  \eque
\equb \times \left( \begin{array}{ccc}
\psi_1\\
\psi_2\\
\psi_3 \end{array} \right) \eque
where the charged lepton mass matrix is already diagonal upon alignment of the $\phi^{\dagger}$ vev in the $(1,1,1)^T$ direction, as in case 23. We therefore want a tribimaximal neutrino mixing matrix once again, now with the  $\psi$ in the $\bold{3_2}$ representation. The invariants in the dimension-5 operator are as in case 25. We can get a tribimaximal neutrino mixing matrix by choosing the following different options for the vev's:
\\

i)
\equb \nonumber  \langle \kappa \rangle = u_1\left( \begin{array}{ccc}
1 \\ 1 \\ 0 \end{array} \right), \,\,\,\,\, \langle \zeta \rangle = u_2\left( \begin{array}{ccc}
0 \\ 1  \\ 1 \end{array} \right) \eque 
where b), c) and k) will contribute.

ii)
\equb \nonumber \langle \chi \rangle = x, \,\,\,\,\, \langle \kappa \rangle = u_1\left( \begin{array}{ccc}
1 \\ 1 \\ 0 \end{array} \right), \,\,\,\,\, \langle \zeta \rangle = u_2\left( \begin{array}{ccc}
0 \\ 1  \\ 1 \end{array} \right) \eque 
leading to the same result as in i).

iii)
\equb \langle \eta \rangle = v_1 \left( \begin{array}{ccc}
1 \\ 0  \\ 0 \end{array} \right), \,\,\,\,\, \langle \kappa \rangle = u_1\left( \begin{array}{ccc}
1 \\ 1 \\ 0 \end{array} \right), \,\,\,\,\, \langle \zeta \rangle = u_2\left( \begin{array}{ccc}
0 \\ 1  \\ 1 \end{array} \right) \eque 
with contributions as in i) plus extra terms from a) and h).

iv)
\equb \langle \rho \rangle = v_2 \left( \begin{array}{ccc}
0 \\ 0  \\ 1 \end{array} \right), \,\,\,\,\, \langle \kappa \rangle = u_1\left( \begin{array}{ccc}
1 \\ 1 \\ 1 \end{array} \right), \,\,\,\,\, \langle \zeta \rangle = u_2\left( \begin{array}{ccc}
0 \\ 1  \\ 1 \end{array} \right) \eque 
with contributions from b), c), i) and k).

v)
\equb  \langle \rho \rangle = v_2 \left( \begin{array}{ccc}
1 \\ 1  \\ 0 \end{array} \right), \,\,\,\,\, \langle \kappa \rangle = u_1\left( \begin{array}{ccc}
1 \\ 1 \\ 1 \end{array} \right), \,\,\,\,\, \langle \zeta \rangle = u_2\left( \begin{array}{ccc}
1 \\ 0  \\ 0 \end{array} \right) \eque 
with terms contributing from b), c), i) and k).
\\

The following corresponding neutrino mass matrices will be possible:
\\

i), ii)
\equb M_{\nu} =  \left( \begin{array}{ccc}
f_2u_1^2  & f'_{19}u_1u_2 &  f'_{19}u_1u_2  \\
 f'_{19}u_1u_2 & f_{3}u_2^2 & f'_4u_2^2 \\
 f'_{19}u_1u_2 & f'_4u_2^2 & f_3u_2^2 \end{array} 
\right) \eque

iii)
\equb M_{\nu} =  \left( \begin{array}{ccc}
f_2u_1^2  +f_{14}v_1u_2 & f'_{19}u_1u_2 &  f'_{19}u_1u_2  \\
 f'_{19}u_1u_2 & f_{3}u_2^2 & f_1v_1^2 + f'_4u_2^2 \\
 f'_{19}u_1u_2 & f_1v_1^2 + f'_4u_2^2 & f_3u_2^2 \end{array} 
\right) \eque

iv)
\equb M_{\nu} =  \left( \begin{array}{ccc}
f_2u_1^2 +f_{17}v_2u_1 & f'_{19}u_1u_2 &  f'_{19}u_1u_2  \\
 f'_{19}u_1u_2 & f_2u_1^2 +  f_{3}u_2^2 & f'_4u_2^2 + f'_{15}v_2u_1\\
 f'_{19}u_1u_2 & f'_4u_2^2 + f'_{15}v_2u_1 & f_2u_1^2 + f_3u_2^2 \end{array} 
\right) \eque

v)
\equb M_{\nu} =  \left( \begin{array}{ccc}
f_2u_1^2 + f_3u_2^2  & f'_{15}v_2u_1 &  f'_{15}v_2u_1  \\
 f'_{15}v_2u_1 & f_{2}u_1^2 + f_{17}v_2u_1 & f'_{19}u_1u_2 \\
f'_{15}v_2u_1 & f'_{19}u_1u_2 &f_{2}u_1^2 + f_{17}v_2u_1 \end{array} 
\right) \eque
all with the same form as in case 23:

\equb M_{\nu} =  \left( \begin{array}{ccc}
\gamma & \beta & \beta  \\
\beta & \alpha & \delta \\
\beta & \delta & \alpha \end{array} 
\right) \eque
This will again give tribimaximal mixing, if $\delta = \gamma + \beta - \alpha$ is fulfilled. 
\\

The selected cases are chosen, because there exists no other result than these. Only the two possible neutrino mass matrices extracted here, can be obtained in this group. We can therefore cover all the cases of the group, by just studying these two forms of the neutrino mass matrix, with the $\psi$ in either a $\bold{3_1}$ or a $\bold{3_2}$ representation.  

\section{Conclusion}

The Frobenius group, with its four 3 dimensional irreducible representations, seems like a suitable group to use, in order to study the family symmetry of neutrinos and charged leptons. 

The approach we have used here is similar to the approach used to study the tetrahedral group in \cite{Tetrahedral}. It is interesting to notice how similar the two groups are with similar irreducible representations, Kronecker products and Clebsch-Gordon decompositions of these, as well as the mass matrices extracted from the Lagrangian. However, the results gained for the vacuum alignments are very different indeed, and the match between the neutrino and charged lepton sector seems to be improved for this Frobenius group. However, to find out which vev is the right to use, we need to place further constraints on the model.

In this paper, we have not addressed the sequestering problem. When the Higgs of the dimension-4 operator acquires a vev in the $(1,1,1)^T$ direction, the $T_{13}$ symmetry is broken down to $Z_3$. As the possible Higgs fields of the dimension-5 operators acquire vev's, some also break $T_{13}$ down to $Z_3$, but in every single case there is at least one vev that breaks it down to $Z_2$. The $Z_2$ subgroup does not commute with the $Z_3$ subgroup, and the vacuum alignments can not be sequestered from each other, unless the interaction terms in the Higgs potential, connecting the fields that break the group down differently, vanishes \cite{Keum}. The tribimaximal mixing is therefore a lowest order approach, in which we need to study allowed deviations within the $T_{13}$ structure. This will impose extra constraints on the vacuum alignments of the Higgs fields, changing the mixing matrix, which thereby might also change the fit between the mass matrices gained from the neutrino and charged lepton sector, respectively. 

Certain articles have reached modifications to the tribimaximal mixing \cite{Minimal} \cite{Harrison}. These could be used for this theory, to find a more constrained match between the two sectors. Also, we have in this paper assumed CP conservation. A violation of this could be considered with an appropriate mixing matrix. 

Furthermore, we have assumed a specific vacuum alignment of the Higgs field in the dimension-4 operator. Other vacuum alignments could be studied. 
\\
\\
$\bold{Acknowledgements}$
\\
\\
We would like to thank the Academia Sinica in Taipei for their great hospitality during the time the ideas for this paper were developed. CH wishes to thank the Augustinus Foundation, the Loerup Foundation and the Reinholdt W. Jorck and Wife's Foundation for funding. AZ acknowledges the support of the NSF under Grant No. PHY07-57035.
\\
\\
$\bold{Note:}$ After our first submission, we have come to learn, that another paper, \cite{Ding}, has carried out studies of the group $T_{13}$ as a model for tribimaximal mixing, using a supersymmetric approach. 

\bibliographystyle{unsrtnat.bst}

\bibliography{Ref_article}

\end{document}